\documentclass[floatfix,lengthcheck,showpacs,amssymb,amsmath,amsfonts,twocolumn]{aastex62}

 \usepackage[utf8]{inputenc}
\usepackage{color}
\usepackage{graphicx}
\usepackage{float}
\usepackage{subfigure}
\usepackage{amsmath}
\usepackage{acronym}
\usepackage{multirow}
\usepackage{tabu}
\usepackage{tikz}
\usetikzlibrary{arrows,shapes,trees,decorations.pathreplacing}
\usepackage{import}
\usepackage{svn-multi}
\usepackage{lineno}
\usepackage{hyperref}
\hypersetup{
    colorlinks=true,
    linkcolor=blue,
    filecolor=magenta,
    urlcolor=cyan,
}

\newcommand{\chieff}{\ensuremath{\chi_{\mathrm{eff}}}}
\newcommand{\rankingstat}{\ensuremath{\tilde{\rho}_c}}
\newcommand{\tdr}{\ensuremath{\mathrm{TDR}}}
\newcommand{\far}{\ensuremath{\mathcal{F}}}
\newcommand{\tar}{\ensuremath{\mathcal{T}}}
\newcommand{\msun}{\ensuremath{\mathrm{M}_{\odot}}}
\newcommand{\pastro}{\ensuremath{P_{\mathrm{astro}}}}
\newcommand{\release}{\texttt{\url{www.github.com/gwastro/1-ogc}}}

\begin{document}
\title[]{1-OGC: The first open gravitational-wave catalog of binary mergers from analysis of public Advanced LIGO data}

\correspondingauthor{Alexander H. Nitz}
\email{alex.nitz@aei.mpg.de}

\author[0000-0002-1850-4587]{Alexander H. Nitz}
\affil{Max-Planck-Institut f{\"u}r Gravitationsphysik (Albert-Einstein-Institut), D-30167 Hannover, Germany}
\affil{Leibniz Universit{\"a}t Hannover, D-30167 Hannover, Germany}

\author[0000-0002-0355-5998]{Collin Capano}
\affil{Max-Planck-Institut f{\"u}r Gravitationsphysik (Albert-Einstein-Institut), D-30167 Hannover, Germany}
\affil{Leibniz Universit{\"a}t Hannover, D-30167 Hannover, Germany}

\author[0000-0001-8694-4026]{Alex B. Nielsen}
\affil{Max-Planck-Institut f{\"u}r Gravitationsphysik (Albert-Einstein-Institut), D-30167 Hannover, Germany}
\affil{Leibniz Universit{\"a}t Hannover, D-30167 Hannover, Germany}

\author[0000-0002-4599-6054]{Steven Reyes}
\affil{Department of Physics, Syracuse University, Syracuse NY 13244, USA}

\author[0000-0002-5192-7784]{Rebecca White}
\affil{Fayetteville-Manlius High School, Manlius, NY 13104, USA}
\affil{Department of Physics, Syracuse University, Syracuse NY 13244, USA}

\author[0000-0002-9180-5765]{Duncan A. Brown}
\affil{Department of Physics, Syracuse University, Syracuse NY 13244, USA}

\author[0000-0003-3015-234X]{Badri Krishnan}
\affil{Max-Planck-Institut f{\"u}r Gravitationsphysik (Albert-Einstein-Institut), D-30167 Hannover, Germany}
\affil{Leibniz Universit{\"a}t Hannover, D-30167 Hannover, Germany}


\begin{abstract}
We present the first Open Gravitational-wave Catalog (1-OGC), obtained by using the public data from Advanced LIGO's first observing run to search for compact-object binary mergers. Our analysis is based on new methods that improve the separation between signals and noise in matched-filter searches for gravitational waves from the merger of compact objects. The three most significant signals in our catalog correspond to the binary black hole mergers GW150914, GW151226, and LVT151012. We assume a common population of binary black holes for these three signals by defining a region of parameter space that is consistent with these events. Under this assumption, we find that LVT151012 has a 97.6\% probability of being astrophysical in origin. No other significant binary black hole candidates are found, nor did we observe any significant binary neutron star or neutron star--black hole candidates. We make available our complete catalog of events, including the sub-threshold population of candidates.
\end{abstract}

\section{Introduction}
\label{sec:intro}

The Advanced LIGO gravitational wave observatories~\citep{Martynov:2016fzi} performed their first observing run (O1) from September 12, 2015 to January 19, 2016. This provided a total of 51.5 days of coincident observations from the two detectors located in Hanford, WA and Livingston, LA. The binary black hole mergers observed in this observing run have been  reported by the LIGO and Virgo Collaborations (LVC) in \cite{Abbott:2016blz,Abbott:2016nmj,TheLIGOScientific:2016pea}.  These binary black hole detections have been independently studied by \cite{Green:2017voq,Roulet:2018jbe,Antelis:2018smo}.

Since the publication of the results by \cite{TheLIGOScientific:2016pea,Abbott:2016ymx}, improvements to the data-analysis methods used~\citep{TheLIGOScientific:2016qqj} have been implemented \citep{Nitz:2017svb,Nitz:2017lco,DalCanton:2017ala}.  Using these improvements, we re-analyze the O1 data and provide---for the first time---a full catalog of candidate events from a matched filter search for compact binary coalescences using the O1 data, which we call 1-OGC. This catalog provides estimates of the significance of previously known events and a ranked list of sub-threshold candidates. Although not significant by themselves, these sub-threshold candidates can be correlated with archival data or transient events found by other astronomical observatories to provide constraints on the population of compact-object mergers \citep{Ashton:2017ykh, Burns:2018pcl}.

Our catalog is based entirely on public, open data and software. We use the LIGO data available from the Gravitational Wave Open Science Center~\citep{Vallisneri:2014vxa}, and analyze the data using the open source PyCBC toolkit~\citep{Usman:2015kfa,Canton:2014ena,pycbc-github}.This toolkit was also used by one of the two analyses described in~\cite{TheLIGOScientific:2016qqj}. The lowest mass sources targeted in our search are neutron star binaries with total mass $m_1 + m_2 = 2\, M_\odot$. The search space extends to binary black hole systems that produce gravitational waveforms longer than $0.15$~s from $20$~Hz. This corresponds to a total mass up to $500 M_{\odot}$ for sources with high mass ratios and spins where the component aligned with the orbital angular momentum is positive and large. For binaries with negligible spin, this corresponds to total mass $\lesssim 200 M_{\odot}$. The search space also includes neutron star--black hole binaries. After applying cuts for data quality~\citep{TheLIGOScientific:2016zmo,TheLIGOScientific:2017lwt}, a total of 48.1~days of coincident data are searched for signals.

The three most significant signals in the catalog correspond to GW150914~\citep{Abbott:2016blz}, LVT151012~\citep{Abbott:2016blz,TheLIGOScientific:2016pea}, and GW151226~\citep{Abbott:2016nmj}, respectively. No other astrophysically significant signals are observed. In the analysis of \cite{TheLIGOScientific:2016pea}, LVT151012 was the third-most significant event, but it was not sufficiently significant to be labeled as an unambiguous detection. With the improved methods employed here, the false alarm rate of this candidate improves by an order of magnitude and it should be considered a true astrophysical event. The analyses of \cite{TheLIGOScientific:2016pea,Abbott:2016ymx} restricted the astrophysical search space to binaries with a total mass less that $100\,M_\odot$. Our analysis extends this target space to higher mass signals. No additional signals are detected in this region of parameter space, consistent with the results of \cite{Abbott:2017iws}.

A second observing run (O2) of the Advanced LIGO detectors took place from November 30, 2016 to August 25, 2017~\citep{Aasi:2013wya}.  The Virgo gravitational wave detector also collected data for part of this period, starting from August 1, 2017.  The detections reported in this second observing run thus far include three additional binary black hole coalescence events \citep{Abbott:2017vtc,Abbott:2017gyy,Abbott:2017oio}, and a binary neutron star merger \citep{TheLIGOScientific:2017qsa}. However, the full O2 data set has not yet been released. The catalog presented here is therefore restricted to the first observing run, O1.

Our paper is organized as follows: In Sec.~\ref{sec:search} and Sec.~\ref{sec:tdr}, we summarize our analysis methods, including the parameter space searched, the detection statistic used for ranking candidate events, and our method for calculating the statistical significance of events. The search results are summarized in Sec.~\ref{sec:results}.  Our full catalog and released data are
described in Sec.~\ref{sec:datarelease} and are available online as supplementary materials (\release). In this paper, we focus on the detection of compact objects. Since no new astrophysical events have been observed, we do not consider measurement of the signals' parameters and refer to \cite{TheLIGOScientific:2016pea,Biwer:2018osg} for discussion of the detected events' source-frame properties. Consequently, we quote binary mass parameters in the detector frame in this work.

\section{Search Methodology}
\label{sec:search}

To search for gravitational waves from compact-object mergers, we use matched filtering~\citep{Allen:2005fk} implemented in the open-source PyCBC library~\citep{Usman:2015kfa,Canton:2014ena,pycbc-github}. Our methods improve on the analyses of \cite{TheLIGOScientific:2016pea,Abbott:2016ymx,TheLIGOScientific:2016qqj} by imposing a phase, amplitude and time delay consistency on candidate signals, an improved background model, and a larger search parameter space~\citep{Nitz:2017svb, Nitz:2017lco, DalCanton:2017ala}.

\begin{figure}[h]
  \centering
    \includegraphics[width=\columnwidth]{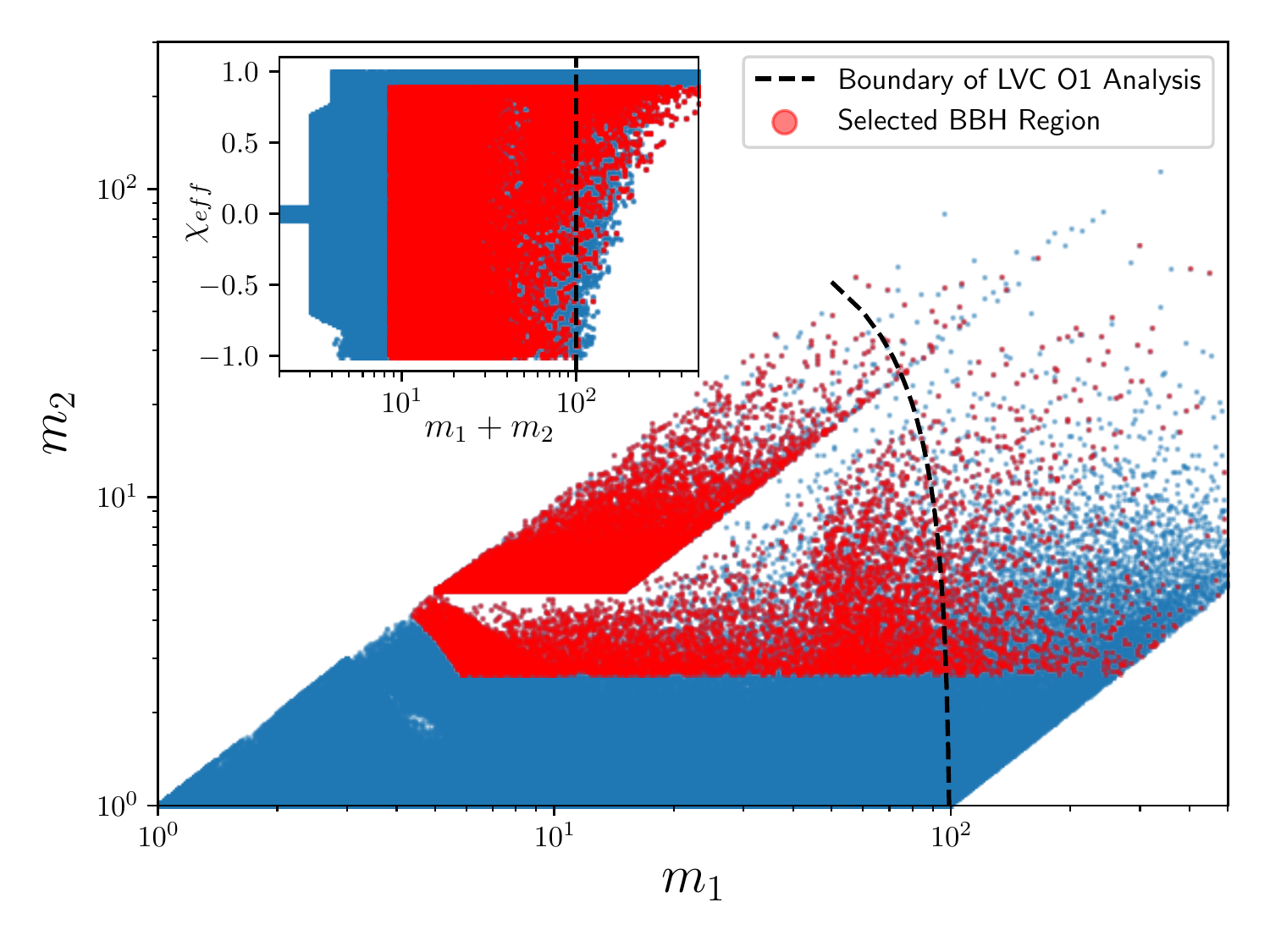}
\caption{The component masses and spins of the templates used to search for compact binary mergers. Due to the exclusion of short duration templates, there is a dependency on the total mass searched and its effective spin. For binary black holes with negligible spin, this implies that this study only probes sources with total mass less than ~$200\,\msun$. Visible artifacts due to the procedure for constructing the template bank do not impact performance. Templates which we conservatively consider to produce binary black hole (BBH) candidates consistent with known observations are shown in red as discussed in Sec.~\ref{sec:tdr}. The upper mass boundary of the analysis performed by the LVC in~\cite{TheLIGOScientific:2016pea} is shown
as a black dotted line.}
\label{fig:bank}
\end{figure}

\subsection{Target Search Space}

A discrete bank of gravitational-wave template waveforms~\citep{Owen:1995tm,Owen:1998dk,Brown:2012qf} is used to target binary neutron star,
neutron star--black hole, and binary black hole mergers with total mass from $2-500 M_{\odot}$~\citep{DalCanton:2017ala}. The templates are
parameterized by their component masses $m_{1,2}$ and their dimensionless spins  $\vec{\chi}_{1,2} = c \vec{S}_{1,2}/G m_{1,2}^2$, where
$\vec{S}_{1,2}$ are the spin vectors of each compact object.  For compact objects with component
masses greater than $2 M_{\odot}$, the template bank covers a wide range of spins, with $\chi_{(1,2)z} \in [\pm 0.998]$, where $\chi_{(1,2)z}$ are the components aligned with the orbital angular momentum. For compact objects
with masses less than  $2 M_{\odot}$, the spin is restricted to $\chi_{(1,2)z} \in [\pm 0.05]$~\citep{Brown:2012qf}. Templates that correspond
to sources with a signal duration less than 0.15 seconds (starting from $20\,$Hz) are excluded due to the difficulty in separating candidates arising from
these templates from populations of instrumental glitches~\citep{DalCanton:2017ala}. Consequently, the total mass boundary of the search
depends strongly on the ``effective spin" \citep{Racine:2008qv, Ajith:2009bn},
\begin{equation}
\chieff = \frac{\chi_{1z} m_1 + \chi_{2z} m_2}{m_1+m_2}.
\end{equation}
This dependence is visible in the distribution of the approximately $400,000$ templates required to cover the space shown in Fig.~\ref{fig:bank}. A dotted line in Fig.~\ref{fig:bank} denotes the upper boundary of the O1 analysis performed in~\cite{TheLIGOScientific:2016pea}. For binaries with total mass greater than $4\,M_\odot$, we use the spinning effective-one-body model (SEOBNRv4)~\citep{Taracchini:2013,Bohe:2016gbl} as template gravitational waveforms. For sources with total masses less than $4M_{\odot}$ we use TaylorF2 post-Newtonian waveforms with phasing accurate to 3.5 post-Newtonian order and the dominant amplitude evolution~\citep{Sathyaprakash:1991mt,Droz:1999qx,Blanchet:2002av,Faye:2012we}. Our choice of template bank discretization causes less than a $10\%$ loss in detection rate for any source within the boundaries of the template bank. Our search assumes that the source can be adequately described by only the dominant gravitational-wave mode, two component masses, non-precessing spins, and negligible eccentricity.

\subsection{Creation and Ranking of Candidate Events}

For each template and each detector, we calculate the matched filter signal-to-noise ratio (SNR) as a function of time $\rho(t)$~\citep{Allen:2005fk}. The template bank is divided into 15 equal sized sub-banks based on the chirp mass $\mathcal{M} = (m_1m_2)^{3/5} / (m_1+m_2)^{1/5}$ of each template. A single-detector ``trigger" is a peak in the SNR time series that is greater than 4 and larger than any other peaks within 1s. For each sub-bank, the loudest 100 triggers (by $\rho$) are recorded in $\sim1$s fixed time windows. This method has been shown to improve search sensitivity, while making the rate of single-detector triggers manageable~\citep{Nitz:2018rgo}. We have found this choice of sub-banks to be an effective method to ensure the analysis can concurrently record triggers from separate regions of parameter space that respond differently to instrumental noise. Other choices are possible.

 We use the data-quality segments provided by the Gravitational-Wave Open Science Center to exclude triggers that occur in times when there are problems with the detectors' data quality~\citep{TheLIGOScientific:2016zmo,TheLIGOScientific:2017lwt}.
 In addition, very loud transient glitches, corresponding to $>100\sigma$ deviations from Gaussian noise, are excised from the strain data according to the procedure of~\cite{Usman:2015kfa} before calculation of the SNR time series. However, there remain many types of transient non-Gaussian noise in the LIGO data which produce triggers with large values of SNR~\citep{Nuttall:2015dqa,TheLIGOScientific:2016zmo,TheLIGOScientific:2017lwt}.

 For every trigger with $\rho > 5.5$ we calculate the signal consistency test, $\chi^2_r$, introduced in~\cite{Allen:2004gu}. The statistic $\chi^2_r$ divides the matched filter into frequency bands and checks that the contribution from each band is consistent with the expected signal. The statistic takes values close to unity when the data contains either Gaussian noise or the expected signal and larger values for many types of transient glitches. We impose the SNR limit as the $\chi^2_r$ test is generally non-informative when $\rho < 5.5$. The $\chi^2_r$ value is used to re-weight the SNR $\rho$ as~\citep{Babak:2012zx}
\begin{equation}
 \tilde{\rho} = \begin{cases}
        \rho & \mathrm{for}\ \chi^2_r \leq 1 \\
        \rho\left[ \frac{1}{2} \left(1 + \left(\chi^2_r\right)^3\right)\right]^{-1/6}, &
        \mathrm{for}\ \chi^2_r > 1.
    \end{cases}
\end{equation}

For single-detector triggers from templates with total mass greater than 40$M_{\odot}$ we apply an additional test, $\chi^2_{r,sg}$, that determines if the detector output contains power at higher frequencies than the maximum expected frequency content of the gravitational-wave signal~\citep{Nitz:2017lco}. This test is only applied for higher mass systems, since these templates are shorter in duration and more difficult to separate from instrumental noise. For other systems, we set $\chi^2_{r,sg} = 1$. Using this statistic, we apply a further re-weighting as
\begin{equation}
\label{eq:sg}
 \hat{\rho} = \begin{cases}
        \tilde{\rho} & \mathrm{for}\ \chi^2_{r,sg} \leq 4 \\
        \tilde{\rho} (\chi^2_{r,sg} / 4)^{-1/2}, &
        \mathrm{for}\ \chi^2_{r,sg} > 4.
    \end{cases}
\end{equation}

Candidate events are generated when single-detector triggers occur in both the LIGO Hanford and Livingston data within $12$~ms (the light-travel time between the observatories extended by $2$~ms for signal time-measurement error) and if the triggers are recorded in the same template in each detector~\citep{Usman:2015kfa}.  Following the procedure of~\cite{Nitz:2017svb}, we model the distribution of single detector triggers from each template as an exponentially decaying function, $\lambda(\hat{\rho}, \vec{\theta}^N)$, where $\vec{\theta}^N$ allows the parameters of the exponential to vary as a function of total mass, symmetric mass ratio $\eta=m_1m_2/M^2$, and $\chieff$. This fitted model allows us to rescale $\hat{\rho}$ to better equalize the rate of triggers from each template.

We improve upon the ranking of candidates in~\cite{Abbott:2016ymx,TheLIGOScientific:2016pea} by also taking into account $p^S(\vec{\theta}^S)$, which is the expected distribution of SNR $\rho_H$ and $\rho_L$, phase difference $\phi_{c, H} - \phi_{c, L}$, and arrival time delay $t_{c,H} - t_{c,L}$ between the two LIGO instruments for an astrophysical population~\citep{Nitz:2017svb}. No assumption is made about the distribution of intrinsic source parameters in this term. The primary benefit arises from assuming the population of sources is isotropically distributed in orientation and sky location. The final ranking statistic \rankingstat{} is then calculated as
\begin{equation}\label{eq:genstat}
  \rankingstat \propto \left[ \log p^S(\vec{\theta}^S) - \log \left(\lambda_H(\hat{\rho}_{H},\vec{\theta}^N) \lambda_L(\hat{\rho}_{L}, \vec{\theta}^N)\right)
  \right] + \mathrm{const.}
\end{equation}
This expression is normalized so that \rankingstat{} approximates the standard network SNR $\rho_c = (\rho_L^2 + \rho_H^2)^{1/2}$ for candidates from regions of parameter space that are not affected by elevated rates of instrumental noise. Candidates from regions affected by elevated rates of noise triggers are down-weighted and assigned a smaller statistic value by this method. As multiple candidates, which arise from different template waveforms, may occur in response to the same signal, we select only the highest ranked candidate within ten seconds. A simpler version of this statistic where the single-detector exponential noise model is only a function of the template duration has also been employed in the analysis of data from LIGO's second observing run~\citep{GW170104, GW170814, Abbott:2017gyy}.

\subsection{Statistical Significance}

The statistical significance of candidate events is estimated by measuring empirically the rate of false alarms (FAR). To measure the noise background rate, we generate additional analyses by time shifting the data from one instrument with respect to the other by multiples of 100 ms. Since this time shift is greater than the maximum astrophysical time of flight between observatories, any candidates produced in these analyses are false alarms. This time shift is much greater than the auto-correlation length of our template waveforms of $\mathcal{O}$(1ms). The time-slid analyses are produced following the same procedure as the search; This is a key requirement for our analysis to produce valid statistical results~\citep{TheLIGOScientific:2016qqj}. The equivalent of more than 50,000 years of observing time can be generated from 5 days of data.

To provide an unbiased measure of the rate of false alarms at least as significant as a potential candidate, the single-detector triggers that compose the candidate event should be included in the background estimation~\citep{2017PhRvD..96h2002C}. However, when a real signal with a large \rankingstat{} is present in the data, the rate of false alarms for candidate events with smaller \rankingstat{} tends to be overestimated. This is due to the fact that the loud single-detector triggers from the real event in one detector form coincidences with noise fluctuations in the other detector, producing loud coincident background events. As in \cite{TheLIGOScientific:2016pea}, an unbiased rate of false alarms can be achieved by a hierarchical procedure whereby a candidate with large \rankingstat{} is removed from the estimation of background for candidates with smaller \rankingstat{}; we use this procedure here.

\section{Evaluating Candidates based on the Astrophysical Population}
\label{sec:tdr}

We find two candidate events with
FAR $< 1$ per $50\,000$ years, corresponding to GW150914 and GW151226.
Although FAR does not give the probability that an event is an astrophysical signal,
we can be confident that these events were not caused by chance
coincidence between the detectors. It is possible that these
events were caused by a correlated source between the detectors. However, detailed followup
studies of GW150914 and GW151226 found no correlated noise sources between the detectors
that could be mistaken for a gravitational wave \citep{TheLIGOScientific:2016zmo, Abbott:2016nmj}.

We conclude that GW150914 and GW151226 are astrophysical in origin and use them to constrain the rate of real signals. A ``true discovery rate"
(\tdr{}) can be constructed for less significant events. The \tdr{} is defined as:
\begin{equation}
\tdr(\rankingstat) = \frac{\tar(\rankingstat)}{\tar(\rankingstat) + \far(\rankingstat)},
\end{equation}
where $\tar(\rankingstat)$ is the rate that signals of astrophysical origin are observed with
a ranking statistic $\geq \rankingstat$ (the ``true alarm rate") and
$\far(\rankingstat)$ is the false alarm rate.

The true discovery rate is the complement of the false discovery rate~\citep{Benjamini:1995ram},
and can be used to estimate the fraction of real signals in a population.
For example, if $\tdr(\rankingstat) = 0.9$, it means that
$90\%$ of events with a ranking statistic $\geq \rankingstat$ are expected to be real signals.  The
\tdr{} is also independent of the observation time.

Note that \tdr{} is not the probability that a particular event is a signal of astrophysical origin \pastro{}. For that, one needs to model the distribution of signals and noise at a given \rankingstat{}. In this work, we use a simple model of these distributions as functions of the ranking statistic \rankingstat{}. Models incorporating additional parameters are also possible, but we do not consider them here. As a function of \rankingstat, \pastro{} can be computed as
\begin{equation}
\pastro(\rankingstat) = \frac{\Lambda_S P_S(\rankingstat)}{\Lambda_S P_S(\rankingstat) + \Lambda_N P_N(\rankingstat)},
\end{equation}
where $P_S(\rankingstat)$ and $P_N(\rankingstat)$ are the probabilities of an event having ranking statistic \rankingstat{}
given the signal and noise hypotheses respectively~\citep{2009MNRAS.396..165G,Farr:2015,Abbott:2016nhf}. $\Lambda_S$ and $\Lambda_N$ are the rates of signal and noise events.

Since no binary neutron star or neutron star--black hole candidates are obtained from a search of the O1 data, here we restrict the calculation of both the \tdr{} and \pastro{} to binary black hole (BBH) observations.
We include signals with total mass $M \geq 10\,\msun$, mass ratio $m_1/m_2 < 5$ (where $m_1 \geq m_2$),
and dimensionless spins $|\chi_{(1,2)z}| <
0.5$. These choices are based on a combination of what has been observed~\citep{TheLIGOScientific:2016pea,GW170104,GW170814,Abbott:2017gyy} and
what is expected from models of isolated binary-star evolution (``field"
binaries). The mass distribution of field binaries is dependent on a number of
unknown parameters, such as the metallicity of the environment~\citep{Belczynski:2014iua}. Generally, it is expected that most binaries are close to equal mass, as typically less than 1
in $\mathcal{O}(10^{3})$ simulated binaries have mass ratio $> 5$ in models of
field-binary evolution \citep{Dominik:2014yma}.  The majority of observations of
nearby X-ray binaries have yielded black holes with masses greater than
$5\,\msun$, which has led to speculation of a ``mass gap" between
3--5$\,\msun$ \citep{Ozel:2010su, Farr:2010tu, Kreidberg:2012ud}. The signals
detected so far by LIGO and Virgo are consistent with this: the smaller component mass
in the lowest-mass system known to date, GW170608, has an estimated mass of
$7^{+2}_{-2}\,\msun$ \citep{Abbott:2017gyy}.

The spin distribution of black holes is not well constrained~\citep{Reynolds:2013qqa}. The component spins
of the most significant binary black holes detected by LIGO and Virgo are
only weakly constrained \citep{TheLIGOScientific:2016pea}. The best measured
quantity related to spin is \chieff{}. All of the BBH gravitational-wave signals
detected so far have $|\chieff| \lesssim 0.2$. A binary with
low \chieff{} may still have component masses with large spin magnitudes,
if the spins are anti-parallel or are purely in the plane of the binary.
However, it seems unlikely that this would be the case for all of the
detections made so far. Hence we include signals that have
component spins with $|\chi_{(1,2)z}| < 0.5$. This is consistent with
recent population synthesis models, which indicate that black holes
must have low natal spin in order to obtain a distribution of \chieff{}
that satisfies gravitational-wave observations \citep{Belczynski:2017gds,Wysocki:2017isg}.

To estimate the rate and distribution of false alarms that arise only
from the region consistent with this selected population of
binary black hole mergers, we must determine which templates are sensitive to these sources.
It is necessary to analyze a simulated set of signals since
the template associated with a particular event is not guaranteed to share the
true source parameters. We find that the region of the template bank defined by
$M > 8.5\,\msun$, $m_{1,2} > 2.7\,\msun$, and $\chieff < 0.9$ is effective at recovering
this population of sources. This region is shown in Fig.~\ref{fig:bank} in red.

To estimate the true rate \tar{}, we use the two significant events observed
during O1, GW150914 and GW151226. We do not use any of the O2 events because the full data is
not yet available for analysis, making it difficult to obtain a consistent rate estimate. The total analysis time in O1 was $\sim48$ days, giving $\tar \approx 15 \mathrm{yr}^{-1}$. Given the uncertainty in
this estimate based on only two events, we take the rate of observations as a Poisson process, and choose the lower
95\% bound on \tar{}. This yields a $\tar \approx 2.7 \mathrm{yr}^{-1}$. For the calculation
of the \tdr{} we use this value for all events, independent of their ranking statistic. This means we likely underestimate the \tdr{} for events quieter than GW151226 and GW150914, but this is a conservative bias.

To estimate the probability that a given event is astrophysical in origin \pastro{}, we model
the distribution of signals and noise as a function of \rankingstat. It is reasonable to approximate the signal
probability distribution $P_S(\rankingstat)$ as $\propto \rankingstat^{-4}$~\citep{Schutz:2011tw,Chen:2014yla}.
We normalize the signal number density $\Lambda_S P_S(\rankingstat)$ so that the number of signals
with $\rankingstat$ greater than or equal to some threshold
$\rankingstat^{\dagger}$ is $\approx 2.7 \mathrm{yr}^{-1}$. We make the conservative choice to place
$\rankingstat^{\dagger}$ at the value of the next largest \rankingstat{} value after GW150914 and GW151226.

\begin{figure}[h]
  \centering
    \includegraphics[width=\columnwidth]{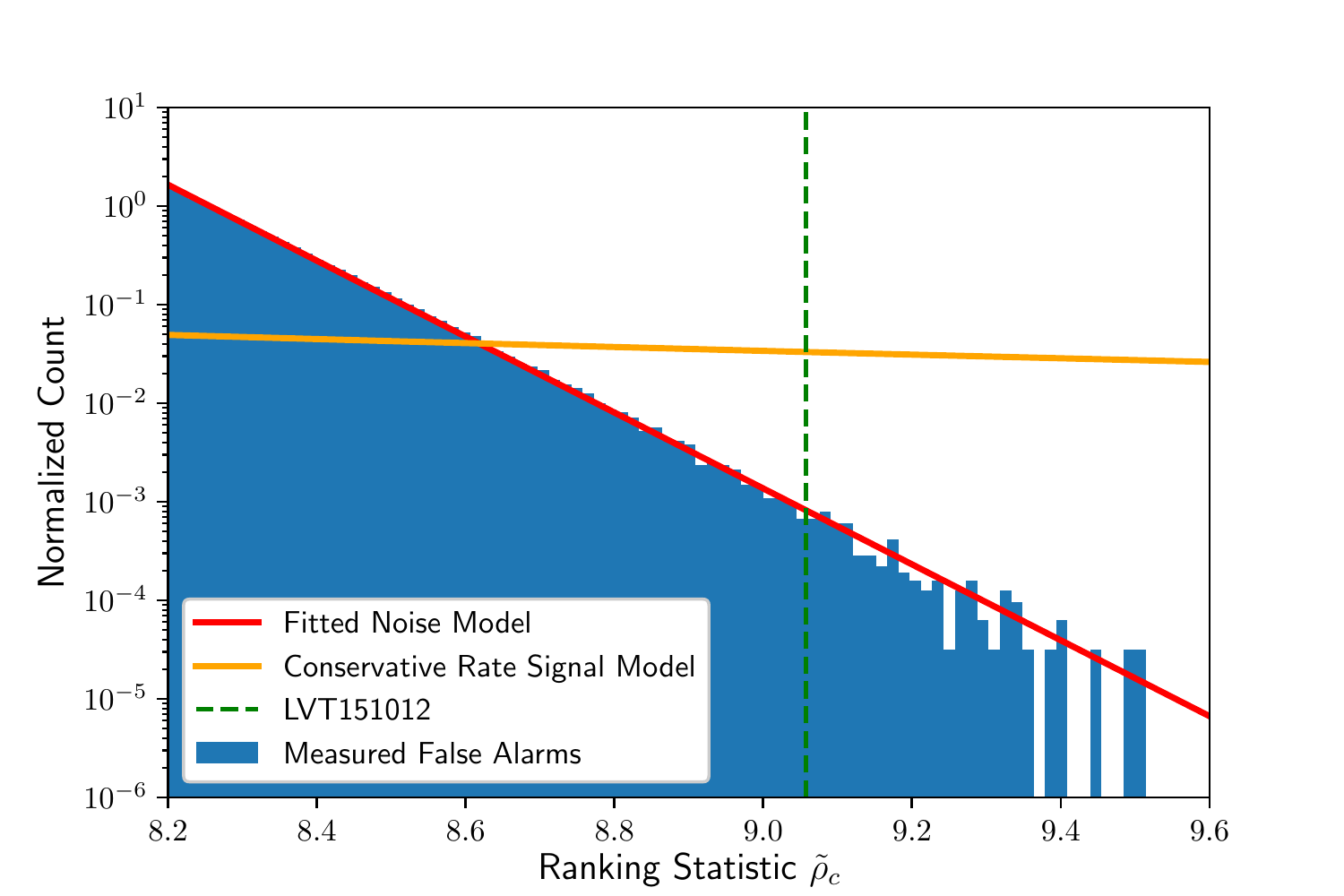}
\caption{The scaled probability distributions of assumed signals and noise as a function of the ranking statistic \rankingstat{} for the analysis containing LVT151012. Blue shows the normalized histogram of empirically measured false alarms that are within our selected BBH region of the template bank, $P_N$. Red is the exponential decay model that has been fitted to this set of false alarms, $P_S \Lambda_S / \Lambda_N$, normalized so that the counts can be directly compared to the noise distribution}. Orange shows the signal model based on our conservative rate of detections. The value of \rankingstat{} for LVT151012 is shown as a dotted green vertical line. The ratio of signal to noise at this value of \rankingstat{} strongly favors the signal model.
\label{fig:pastro}
\end{figure}

To approximate the noise number density $\Lambda_N P_N(\rankingstat)$, we make a histogram of the \rankingstat{}
values of false alarms arising from our selected BBH region. We use only the false alarms which are uncorrelated
with possible candidate events to ensure an unbiased estimate of the mean false alarm rate~\citep{2017PhRvD..96h2002C}.
We fit an exponential decay to this histogram from $8<\rankingstat<9.2$. For \rankingstat{} much less than $8$,
$\Lambda_N P_N$ is not well modeled by an exponential due to the effects of applying a threshold to single-detector
triggers. We note, however, there is only a $ 50 \%$ chance that an event is astrophysical at
\rankingstat{} $\sim 8.6$, and this chance quickly becomes negligible with decreasing \rankingstat. The result of this
procedure is shown in Fig.~\ref{fig:pastro}. We caution that \pastro{} for candidates with
\rankingstat{} $>9.2$ will be sensitive to the form of the model chosen since it is not constrained by
empirically measured false alarms.

While we do not assess the astrophysical probabilities of sources outside our selected BBH region, we are not precluding that such sources exist.
Our \pastro{} is compatible with any model of the true BBH
source distribution that allows for a signal rate to be at least as high as our estimate within the chosen region. This holds irrespective of whatever other kinds of sources may also be permitted.

\begin{figure}[]
  \centering
    \includegraphics[width=\columnwidth]{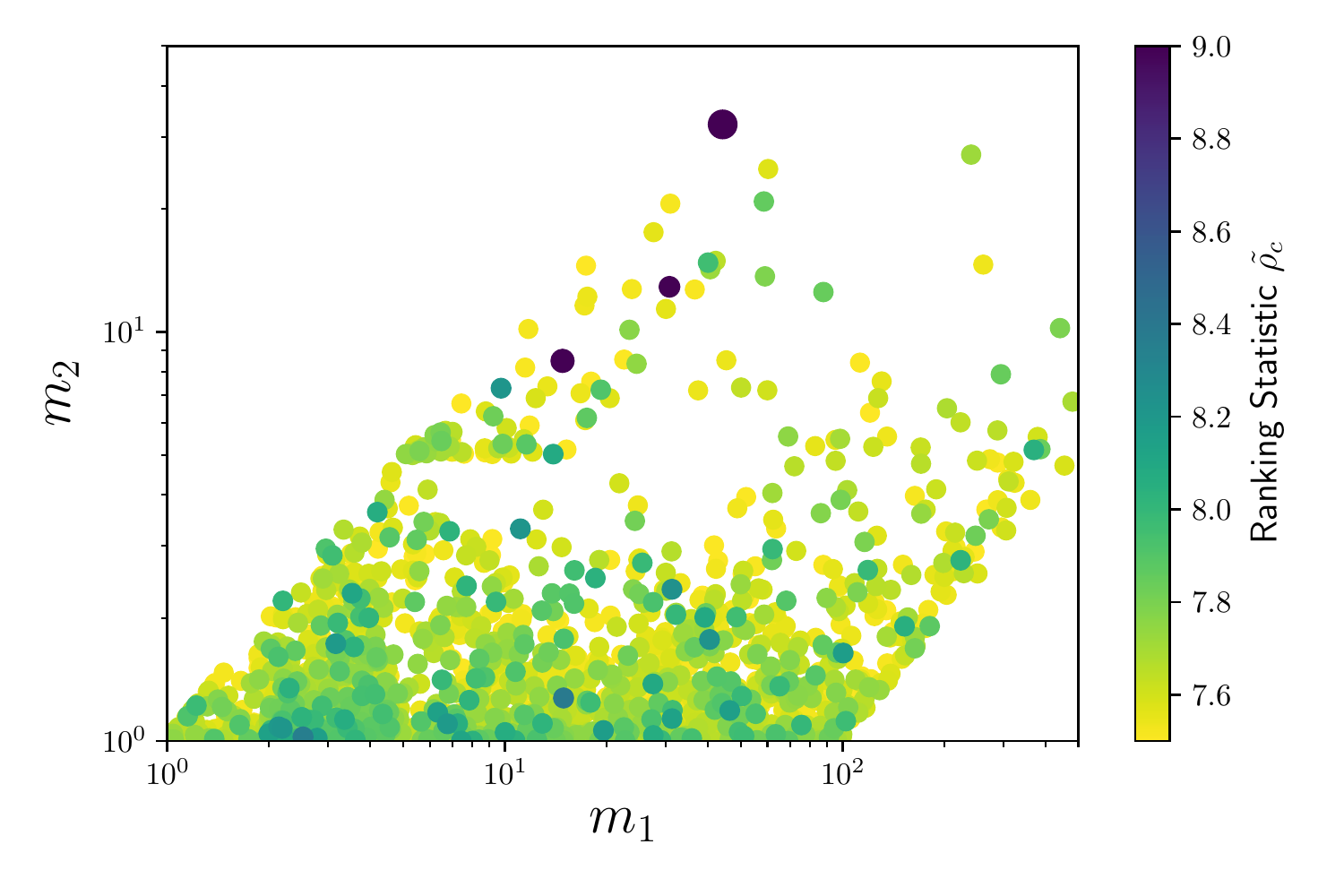}
\caption{Candidate events with a ranking statistic $\rankingstat>7.5$ from the full search for compact binary mergers in O1 data. The colorbar is capped at 9. The three BBH mergers are clearly visible in the plots, while the remaining events are largely distributed according to the density
of the template bank.}
\label{fig:bankcandidates}
\end{figure}

\begin{table*}[]
  \begin{center}
    \caption{Candidate events from the full search for compact binary mergers in O1 data. Candidates are sorted by FAR evaluated for the entire bank of templates. The FAR of the top two candidates is limited only by the amount of background time estimated, and only differ due to the variation in time available in their respective analyses to create background. The parameters of the template associated with each candidate are listed. Note that these are not intended as a rigorous estimation of the source parameters. Masses are given in the detector frame.}
    \label{table:complete}
\begin{tabular}{lllllllll}
Designation & Julian Date & $FAR^{-1} (yr)$ & \rankingstat{} & $\rho_H$ & $\rho_L$ & $m_1$ & $m_2$ & $\chieff{}$ \\ \hline
150914+09:50:45UTC & 2457279.910665 &  $>$66000 & 18.45 & 19.67 & 13.38 & 44.21 & 32.16 & 0.09 \\
151226+03:38:53UTC & 2457382.652426 &  $>$59000 & 11.62 & 10.73 & 7.43 & 14.83 & 8.50 & 0.24 \\
151012+09:54:43UTC & 2457307.913420 &     24 & 9.06 & 6.96 & 6.71 & 30.75 & 12.89 & -0.05 \\
151019+00:23:16UTC & 2457314.516585 &      0.060 & 8.39 & 6.81 & 5.47 & 14.93 & 1.27 & 0.11 \\
150928+10:49:00UTC & 2457293.951122 &      0.042 & 8.37 & 6.05 & 6.34 & 2.53 & 1.02 & -0.70 \\
151218+18:30:58UTC & 2457375.271929 &      0.029 & 8.24 & 7.11 & 5.38 & 31.29 & 2.35 & -0.00 \\
160103+05:48:36UTC & 2457390.742504 &      0.026 & 8.22 & 6.01 & 6.60 & 9.75 & 7.29 & 0.49 \\
151202+01:18:13UTC & 2457358.554740 &      0.025 & 8.23 & 6.54 & 5.73 & 40.42 & 1.77 & -0.26 \\
160104+03:51:51UTC & 2457391.661424 &      0.021 & 8.19 & 5.80 & 6.39 & 6.76 & 1.10 & -0.51 \\
151213+00:12:20UTC & 2457369.508985 &      0.019 & 8.22 & 5.70 & 7.24 & 11.12 & 3.30 & -0.79 \\
150923+07:10:59UTC & 2457288.799711 &      0.014 & 8.20 & 6.78 & 5.84 & 2.14 & 1.08 & 0.65 \\
151029+13:34:39UTC & 2457325.066149 &      0.014 & 8.21 & 6.83 & 5.23 & 2.19 & 1.07 & -0.27 \\
151206+14:19:29UTC & 2457363.097291 &      0.013 & 8.17 & 5.80 & 6.37 & 100.60 & 1.64 & 0.98 \\
151202+15:32:09UTC & 2457359.147751 &      0.012 & 8.14 & 5.93 & 6.41 & 6.33 & 1.18 & -0.59 \\
151012+06:30:45UTC & 2457307.771774 &      0.011 & 8.19 & 6.74 & 5.70 & 3.16 & 1.73 & -0.15 \\
151116+22:41:48UTC & 2457343.446120 &      0.010 & 8.14 & 5.79 & 6.64 & 2.00 & 1.04 & -0.45 \\
151121+03:34:09UTC & 2457347.649138 &      0.010 & 8.12 & 6.48 & 5.78 & 7.43 & 1.00 & -0.86 \\
150922+05:41:08UTC & 2457287.737317 &      0.010 & 8.16 & 6.05 & 6.34 & 2.78 & 1.02 & 0.17 \\
151008+14:09:17UTC & 2457304.090202 &      0.008 & 8.16 & 5.84 & 6.10 & 46.38 & 1.19 & 0.38 \\
151127+02:00:30UTC & 2457353.584101 &      0.008 & 8.10 & 6.28 & 5.44 & 39.12 & 2.01 & 0.99 \\

\end{tabular}
  \end{center}
\end{table*}

\section{Results}
\label{sec:results}

The results presented here are generated using the data from the first observing run of Advanced LIGO which ran from September 12, 2015 to January 19, 2016. We divide the 16~kHz LIGO open data into 9 consecutive periods of time and search each time period independently so that each analysis contains roughly five days of observing time. This time interval is set by the disk and memory requirements of the search pipeline, but it is sufficient to estimate the FAR of candidate events to better than 1 in 50,000 years. It is possible to combine these time intervals during the analysis to improve this limit, but we have not done so here. Our analysis is restricted to times marked as observable by the metadata provided by the Gravitational-Wave Open Science Center. After accounting for times which are marked as not analyzable, there remain $\sim48.1$ days of data when both the Hanford and Livingston LIGO instruments were operating.

The top candidate events by FAR from the full search are given in Table~\ref{table:complete}. There are three candidates which are statistically significant. These are the binary black hole mergers GW150914, LVT151012, and GW151226, which were previously reported in~\cite{TheLIGOScientific:2016pea,Abbott:2016blz,Abbott:2016nmj}. The false alarm rates for GW150914 and GW151226 of 1 per 66,000 and 1 per 59,000 years, respectively, are limits based on the amount of background time available in their respective analysis. These limits are less stringent than those reported in~\cite{TheLIGOScientific:2016pea} as we have created less background time. There are no other individually convincing candidates. Fig.~\ref{fig:bankcandidates} shows candidate events with $\rankingstat > 7.5$. The three binary black hole mergers stand out from the other candidate events and are clustered in a portion of the parameter space that is analyzed with relatively few template waveforms.

\begin{table*}[]
  \begin{center}
    \caption{Candidate events consistent with the selected population of binary black holes. There are three binary black hole mergers above a threshold corresponding to a true discovery rate of $99.92\%$. The third most significant event, LVT151012, has a 97.6\% probability of
    being astrophysical in origin. Note that the FARs indicated do not reflect the false alarm rate for the full search, but instead for the limited region of the template bank indicated in red in Fig.~\ref{fig:bank}. The FARs listed for the top two events are limited
    by the background time generated and so are identical to those in Table~\ref{table:complete}.}
    \label{table:bbh}
\begin{tabular}{lllllllllll}
Designation & Julian Date & $\pastro{}$ & TDR & $FAR^{-1} (yr)$ & \rankingstat{} & $\rho_H$ & $\rho_L$ & $m_1$ & $m_2$ & $\chieff{}$ \\ \hline
150914+09:50:45UTC & 2457279.910665 & - & - &  $>$66000 & 18.45 & 19.67 & 13.38 & 44.21 & 32.16 & 0.09\\
151226+03:38:53UTC & 2457382.652426 & - & - &  $>$59000 & 11.62 & 10.73 & 7.43 & 14.83 & 8.50 & 0.24\\
151012+09:54:43UTC & 2457307.913420 & 0.976 & 0.999 &    446 & 9.06 & 6.96 & 6.71 & 30.75 & 12.89 & -0.05\\
160103+05:48:36UTC & 2457390.742504 & 0.061 & 0.517 &      0.396 & 8.22 & 6.01 & 6.60 & 9.75 & 7.29 & 0.49\\
151213+00:12:20UTC & 2457369.508985 & 0.047 & 0.455 &      0.309 & 8.22 & 5.70 & 7.24 & 11.12 & 3.30 & -0.79\\
151216+18:49:30UTC & 2457373.284799 & 0.017 & 0.223 &      0.106 & 8.09 & 6.10 & 6.01 & 13.92 & 5.03 & -0.41\\
151222+05:28:26UTC & 2457378.728506 & 0.012 & 0.169 &      0.075 & 8.03 & 5.67 & 6.46 & 6.86 & 3.26 & -0.74\\
151217+03:47:49UTC & 2457373.658627 & 0.006 & 0.088 &      0.036 & 7.96 & 6.69 & 5.57 & 40.02 & 14.77 & 0.84\\
151009+05:06:12UTC & 2457304.713060 & 0.005 & 0.087 &      0.035 & 7.99 & 5.66 & 5.90 & 25.55 & 2.73 & -0.05\\
151220+07:45:36UTC & 2457376.823761 & 0.003 & 0.053 &      0.021 & 7.87 & 6.55 & 5.39 & 17.50 & 6.17 & 0.82\\
151104+04:12:55UTC & 2457330.676062 & 0.003 & 0.053 &      0.021 & 7.91 & 5.94 & 6.33 & 19.25 & 7.22 & 0.71\\
151120+16:20:06UTC & 2457347.181049 & 0.003 & 0.047 &      0.018 & 7.86 & 6.11 & 5.44 & 5.49 & 3.10 & 0.79\\
151216+09:24:16UTC & 2457372.892271 & 0.003 & 0.045 &      0.017 & 7.86 & 5.76 & 5.66 & 58.56 & 20.84 & 0.66\\
151128+14:37:02UTC & 2457355.109478 & 0.003 & 0.040 &      0.016 & 7.83 & 6.79 & 5.02 & 9.25 & 6.22 & -0.87\\
160109+08:08:42UTC & 2457396.839798 & 0.003 & 0.035 &      0.014 & 7.82 & 5.24 & 6.23 & 24.29 & 3.45 & -0.98\\
160111+22:49:34UTC & 2457399.451507 & 0.003 & 0.035 &      0.013 & 7.82 & 5.10 & 6.55 & 5.75 & 3.43 & 0.23\\
151124+11:25:19UTC & 2457350.976339 & 0.002 & 0.033 &      0.013 & 7.81 & 5.65 & 6.27 & 98.89 & 3.89 & 0.45\\
150912+15:39:02UTC & 2457278.152523 & 0.002 & 0.032 &      0.012 & 7.84 & 6.23 & 5.23 & 9.86 & 5.33 & -0.01\\
151006+06:06:50UTC & 2457301.755168 & 0.002 & 0.031 &      0.012 & 7.89 & 6.77 & 5.47 & 11.59 & 5.31 & -0.05\\
151015+01:40:52UTC & 2457310.570466 & 0.002 & 0.029 &      0.011 & 7.85 & 5.37 & 5.92 & 87.87 & 12.52 & 0.75\\

\end{tabular}
  \end{center}
\end{table*}

\subsection{Binary Black Hole Candidates}

Given that there are two binary black hole mergers (GW150914 and GW151226 ) that are well established from their
statistical significance, we can estimate the rate of detecting binary black hole mergers by this analysis. Candidate
events that are consistent with our selected binary black hole population are listed in
Table~\ref{table:bbh}. We estimate the false alarm rate of events for just this region of the analysis, and using our
estimate of the true rate of detections, calculate the true discovery rate as a function of ranking statistic. The
\tdr{} at the ranking statistic of the fourth most significant candidate is 0.52. This means that only 52\% of
candidates with \rankingstat{} at least as large are expected to be of astrophysical origin. For each candidate we estimate its
individual probability of being astrophysical in origin, \pastro{}. The fourth event has only a 6$\%$ chance of being
astrophysical. We do not report \pastro{} and \tdr{} values for the top two events since these events
are assumed to be signals in the construction of these statistics.

\subsection{Revisiting LVT151012}

LVT151012 was first announced in~\cite{TheLIGOScientific:2016qqj}, with a FAR
of 1 per 2.3 years. Our improved methods yield a false alarm rate for LVT151012
of 1 per 24 years. Restricting attention to our selected BBH region, which is
consistent with the other observed binary black hole mergers, gives a FAR for
LVT151012 in this region alone of 1 per 446 years. We combine this FAR  with
our conservative estimate of the rate of detections to estimate that 99.92\% of
binary black hole merger candidates at least as significant as LVT151012 are
astrophysical in origin. We also estimate the probability that specifically
LVT151012 is astrophysical in origin to be 97.59$\%$.

These measures both depend on our selected region of binary black hole sources
and our estimate of the rate of true detections, but we believe our choices for
both of these to be conservative. The FAR of 1 per 446 years is not a
statistical statement about the search as a whole and is used only in
comparison against the rate of real signals within this same region. Selecting
different boundaries for this region would yield a different FAR. However,
assuming that the false alarm rate and true alarm rate are both approximately
uniform in this region, then \pastro{} and \tdr{} will not change.

As data from future observing runs becomes available, it will be possible to more precisely estimate this rate in a consistent way, and improve our estimate of this event's significance.  We have modeled our signal distribution and population of false alarms as being characterized by the ranking statistic \rankingstat{} alone. An improved model could take into account the variation over the parameter space and in time. Fig.~\ref{fig:pastro} shows the probability distribution of our noise and signal models for the analysis which contains LVT151012. Compared to the \pastro{} reported in~\cite{TheLIGOScientific:2016pea} of 87\%, our analysis has improved the ranking of candidate events, the boundaries of our selected BBH distribution differ from what was used there, and we use a more conservative estimate of the signal rate. Given a \pastro{} value of 97.6$\%$ we conclude that LVT151012 is astrophysical in origin. For comparison, if we had chosen the rate of observed mergers to be $\approx 15 \mathrm{yr}^{-1}$, which is the linear extrapolation of two detections in 48 days, we would find that LVT151012 had a $99.6\%$ probability of astrophysical origin.

\section{Data Release}
\label{sec:datarelease}

The 1-OGC catalog contains $\sim 150,000$ candidate events. Our supplemental materials online provide the complete combined set of binary neutron star, neutron star--black hole, and binary black hole candidates~\citep{1-OGC}. A separate listing of the candidates from our selected BBH region is also made available. Each candidate is assigned an identifying name constructed from the date and UTC time. The vast majority of these candidates are not astrophysical in origin. To help distinguish between possible sources we provide our ranking statistic $\rankingstat$ along with our estimate of the false alarm rate for each candidate. We also provide information such as the SNR observed by each instrument, the time of arrival, measured phases, and the results of our set of signal consistency tests. The periods of time that were analyzed are also provided. We also provide the PyCBC pipeline configuration files that allow our analysis to be reproduced.

\section{Discussion}

We present a full catalog of gravitational-wave events and candidates from a PyCBC-based, templated, matched-filter search of the LIGO O1 open data.  Our analysis represents an improvement over that of \cite{TheLIGOScientific:2016pea,Abbott:2016ymx} by using improved ranking of candidates by considering phase, amplitude and time delay consistency, an improved background model and a template bank targeting a wider range of sources \citep{Nitz:2017svb, Nitz:2017lco,DalCanton:2017ala}. We independently verify the discovery of GW150914 and GW151226 and report an improved significance of the candidate event LVT151012, which we claim should be viewed as a confident detection.  Apart from these three signals, none of the other candidate events are individually significant in our analysis.  All of these candidates are listed in our catalog available at \release{}, along with tools for exploring and using it. Complete gravitational-wave event catalogs of this nature will become important tools in multi-messenger astronomy.

A  larger data set from the second observing run of LIGO and Virgo already exists. Individual detections have been published, and short periods of data around the detections are available publicly.  However, the bulk of this data has not yet been released publicly. It will be possible to create a similar open catalog with the most up-to-date analysis tools when these data are released.

\acknowledgments
We thank Thomas Dent and Sumit Kumar for useful discussions and comments. We thank Stuart Anderson, Jonah Kannah, and Alan Weinstein for help accessing data from the Gravitational-Wave Open Science Center.  We acknowledge the Max Planck Gesellschaft for support and the Atlas cluster computing team at AEI Hannover. Computations were also supported by Syracuse University and NSF award OAC-1541396. DAB acknowledges NSF awards PHY-1707954, OAC-1443047, and OAC-1738962 for support. SR acknowledges NSF award PHY-1707954 and OAC-1443047 for support. RW acknowledges NSF award OAC-1823378 for support.
This research has made use of data, software and/or web tools obtained from the Gravitational Wave Open Science Center (https://www.gw-openscience.org), a service of LIGO Laboratory, the LIGO Scientific Collaboration and the Virgo Collaboration. LIGO is funded by the U.S. National Science Foundation. Virgo is funded by the French Centre National de Recherche Scientifique (CNRS), the Italian Istituto Nazionale della Fisica Nucleare (INFN) and the Dutch Nikhef, with contributions by Polish and Hungarian institutes.
\bibliography{references}

\end{document}